\documentclass[nofootinbib,12pt]{revtex4} 
\usepackage{graphicx}
\usepackage{cancel}
\usepackage{amssymb}
\usepackage{textcomp}
\def\llpp{\ell^+\ell^+}
\def\llmm{\ell^-\ell^-}
\def\llpm{\ell^\pm\ell^\pm}
\def\llij{\ell^+_i \ell^+_j}
\def\HH{H^{++}H^{--}}
\def\WW{W^{\pm}W^{\pm}}
\def\yll{Y_{\ell\ell}}
\def\lsim{\mathrel{\raise.3ex\hbox{$<$\kern-.75em\lower1ex\hbox{$\sim$}}}}
\def\gsim{\mathrel{\raise.3ex\hbox{$>$\kern-.75em\lower1ex\hbox{$\sim$}}}}
\begin{document}

\begin{titlepage}
\begin{flushright}
MADPH-07-1488\\
HRI-P-07-05-001
\end{flushright}
\vspace{0.5cm}
\title{Pair Production of Doubly-Charged Scalars:\\
 Neutrino Mass Constraints and Signals at the LHC}

\author{Tao Han~$^{\bf a,d}$}
\email{than@hep.wisc.edu}

\author{Biswarup Mukhopadhyaya~$^{\bf b}$}
\email{biswarup@mri.ernet.in}

\author{Zongguo Si~$^{\bf c}$}
\email{zgsi@sdu.edu.cn}

\author{Kai Wang~$^{\bf a}$}
\email{wangkai@hep.wisc.edu}

\affiliation{$^{\bf a}$ Department of Physics, University of Wisconsin, Madison, WI 53706, USA\\
$^{\bf b}$ Harish-Chandra Research Institute, Allahabad 211019, INDIA\\
$^{\bf c}$  Department of Physics, Shandong University, Jinan, Shandong 250100, P.R.~China\\
$^{\bf d}$  Center for High Energy Physics, Tsinghua University, Beijing 100084, P.R.~China}

\begin{abstract}
We study the pair production of doubly charged Higgs bosons 
at the Large Hadron Collider (LHC),  assuming the doubly charged Higgs to 
be part of an SU(2)$_L$ triplet which generates Majorana
masses for left-handed neutrinos. Such pair-production
has the advantage that it is not constrained by the triplet vacuum
expectation value, which tends to make the single production rate 
rather small. We point out that, 
in addition to the Drell-Yan (DY) production mechanism,  
two-photon processes also contribute to $\HH$ production 
at a level comparable to the QCD corrections to the DY channel.
Decays of the doubly charged Higgs into both the $\llpp$
and $W^{+}W^{+}$ modes are studied in detail to optimize the signal observation
over the backgrounds. Doubly charged scalars  should be observable at the LHC
with 300 $\rm{fb}^{-1}$ integrated luminosity in the $\llpm$ channel upto the 
mass range of 1 TeV even with a branching fraction of about $60\%$, 
and in the $\WW$ channel upto a mass of 700 GeV. 
Such a doubly charged Higgs,
if  it is  a member  of a triplet generating neutrino masses, 
cannot be long-lived on the scale of collider detectors although it might lead to
a displaced secondary vertex during its decay if it is lighter than about 250 GeV. 
\end{abstract}

\maketitle
\end{titlepage}

\section{Introduction}
Higgs bosons in representations of SU(2)$_L$ other than doublets occur in many extensions
of the standard model.  Among these,  situations of special interest are created by 
SU(2)$_L$ triplet scalars  which occur in various scenarios, ranging from left-right symmetric
models to Little Higgs 
theories \cite{Senjanovic:1975rk,Georgi:1985nv,Gunion:1989in,Arkani-Hamed:2001nc}.  
Complex scalar triplets with a hypercharge $Y = 2$ are particularly rich in this
context in terms of their phenomenological implications.  
Once we allow lepton number violation
by two units, the complex  triplet can couple to 
left-handed leptons in a gauge-invariant 
and renormalizable manner 
and give rise to Majorana masses for neutrinos.  No right-handed 
neutrinos need to be postulated in such a scenario. 
At the same time,  the complex triplet contains a doubly 
charged component ($H^{++}$), 
and the same $\Delta L = 2$  interactions with charged leptons open 
up a very spectacular set of decay channels 
for  this state, namely, resonant decays into  a pair of like-sign  leptons.  
These channels not only lead to
remarkably background-free signatures of the doubly charged scalars, but 
also demonstrate a crucial
link between observations at high energy colliders and the 
widely discussed mechanism of neutrino
mass generation. 
We explore such a link in this paper, 
in the context of the Large Hadron Collider (LHC).   

Although the general idea is simple here and has been 
discussed earlier in different 
connections, including direct experimental searches  at the
Fermilab Tevatron \cite{Acosta:2004uj,Acosta:2005np,Abazov:2004au},  
a number of subtle and challenging issues  
invariably come up in such a study,
and we have done our best to address them.

First of all, the
presence of a $Y=2$ triplet $H~=~(H^{++},H^{+},H^{0})$  
allows the following $\Delta L = 2$  interaction with left-handed lepton
doublets \cite{Ma:2000xh,Han:2005nk}:
\begin{equation}
	{\cal L} = \yll^{ij} L_i^T ~H ~ C^{-1} L_j + {\rm h.c.}
	 \label{lphil}
\end{equation}
where $C$ is the charge conjugation operator. 
In general,  the triplet can develop a vacuum expectation value (vev, $v'$).
It thus contributes to  $\Delta L=2$ Majorana masses for the neutrinos,
proportional to $\yll v'$. The current observations from the 
neutrino oscillation experiments and cosmological bounds yield \cite{pdg} 
\begin{equation}
m_\nu = \yll\  v' \lsim 10^{-10}\ {\rm GeV}.
\label{yvev}
\end{equation}
This implies a stringent bound on $v'$. For instance, if we take $\yll$
to be as small as the electron's Yukawa coupling, 
we have $v'\sim 10^{-2}$ MeV. As for any $\Delta L=2$ interactions,  
neutrinoless double beta decay experiments ($0\nu\beta\beta$) \cite{doublebeta}
provide a direct test. $H^{++}$ also contributes to $0\nu\beta\beta$ via
$W^- W^- \rightarrow H^{--} \rightarrow \ell^- \ell^- .$ 
The $W^-W^-H^{++}$ coupling is proportional to $ v'$ in the model under discussion,  
and with the  $0\nu\beta\beta$ bound  
\cite{Belanger:1995nh,Hirsch:1996qw,Aalseth:2004hb}, we have
\begin{equation}
\frac{v' \yll}{m^2_{H^{++}}} \leq 5\times 10^{-8}~{\rm GeV}^{-1}.
\end{equation}
Given the neutrino mass requirement in Eq.~(\ref{yvev}), 
we obtain a bound $m_{H^{++}} > 0.1~{\rm GeV}$, which is too weak 
to be relevant here.

In fact, there is another direct bound on $v'$ due to the 
electroweak $\rho$-parameter.
In order to prevent large tree-level contributions 
to the $\rho$-parameter, one
needs  \cite{pdg,Gunion:1990dt}
$ v' \lsim 1\ {\rm GeV}. $
Although models exist in the literature \cite{Georgi:1985nv,Gunion:1989ci,Gunion:1990dt}, 
where one has complex as well as real triplets, whose
combined contributions to the $\rho$-parameter 
cancel, thus allowing large triplet vevs
at the tree level,  they presuppose the existence of additional symmetries 
whose validity is not clear once higher order 
effects involving gauge couplings are included. 

It is also to be noted that, if $v'$ is allowed to be close to the upper limit
from the $\rho$-parameter, then the couplings  $Y_{\ell\ell}$ are
forced to be $\simeq 10^{-10}$. Since there are six of these couplings
(assuming the matrix to be real symmetric), 
they all need to be adjusted within
such a small range in order to reproduce the neutrino mass matrix that
fits the observed mixing pattern.  This makes the scenario rather more
fine-tuned than one in which one has a much smaller $v'$ a single quantity), 
 along with the six parameters  $Y_{\ell\ell}$  which are allowed to be
closer to unity.  In this phenomenological study, we will assume a small $v'$
and be guided for the parameters by the neutrino mass generation, to 
saturate the relation of Eq.~(\ref{yvev}).

Secondly, we wish to identify the leading production 
channels for the doubly charged Higgs at the LHC.
%
Rather encouraging predictions about single production of the doubly charged Higgs boson 
in $W$-boson fusion can be found in some earlier studies via the process 
$W^+W^+ \to H^{++}$ \cite{dicus,huitu,atlasdc,Chivukula:1986sp,Gunion:1989ci,Gunion:1990dt}.
However, specific features of models are often utilized in such studies, and
it is difficult to maintain such optimism in a 
general case if the triplet vev is restricted to a smaller value. 
Numerically, the signal rate is proportional to $(v'/v)^2$ and 
is already too small to be observable at the LHC if $v' \lsim 1$ GeV. 
Since we are considering even smaller values of $v'$,
we therefore concentrate instead on pair production 
which is largely governed by electromagnetic interactions.
In principle, the production of one 
doubly-charged Higgs in conjunction with a singly charged one, 
followed by the decay $H^{++} \longrightarrow H^+ W^+$, driven by the
$SU(2)_L$ gauge coupling, can generate additional signals \cite{akeroyd}. 
However, this presupposes considerable mass separation between the 
$H^{++}$ and the  $H^{+}$, which may not be quite expected in many 
scenarios such as those based on Little Higgs theories. Therefore, we 
concentrate on the pair production process as the constantly available 
fall-back, and  remember that the doubly charged scalars produced most 
copiously in the Drell-Yan (DY) channel. In our analysis, we also 
include the production via the two-photon fusion channel. This is motivated 
by the stronger electromagnetic coupling of a doubly charged
particle. Comparing with a singly charged scalar, the two-photon channel will
have an enhancement factor of 16. Numerically, it 
provides about 10 per cent  correction to the DY process, 
and it is comparable to the QCD corrections  \cite{Muhlleitner:2003me}.


Thirdly, as for the identification of the $H^{\pm\pm}$ signal, 
one must consider
both $H^{++} \to \llpp $  and  $H^{++} \to W^+ W^+$,
keeping in mind the interplay of the two independent 
parameters $\yll$ and $v'$  that govern the two decay channels. 
In spite of the simple and distinctive nature of the first
signal,  one cannot
rule out their faking by a number of sources. 
We have studied in detail the event selection 
criteria which establish the {\it bona fide} of such signals 
in a convincing way. Side by side, we are also
suggesting ways of isolating the signals coming from
$H^{++} \to W^+ W^+$, which are in general more background-prone,
and can dominate the decays of the doubly charged scalar in
certain regions of the parameter space.
Most studies on the signals of doubly charged scalars have not
adequately addressed the challenges 
posed by backgrounds in this search channel. 
Furthermore, given the small values of $\yll  v'$,  it is useful 
to investigate whether the 
doubly charged Higgs can indeed be long-lived on the scale 
of collider detectors  
in any region of the parameter space of this scenario,  a situation 
experimentalists have already looked for 
at the Tevatron \cite{Acosta:2005np}. 
And finally, we wish to point out,  albeit in a qualitative way,  
that relative strengths of the 
different flavor diagonal as well as 
off-diagonal decays  $H^{++} \to \llij$   of the doubly 
charged state should give us
information about the matrix  $\yll$ and, in turn, generate 
insight into the structure of the neutrino
mass matrix \cite{Ma:2000xh}, which is responsible for the tri-bimaximal 
mixing pattern suggested by observations.

We present the calculation related to both 
the Drell-Yan and two-photon production 
channels at the LHC in Sec.~II. Section III contains a discussion 
on the decay of the doubly charged states 
into the $\ell^{\pm}\ell^{\pm}$  and $W^{\pm}W^{\pm}$ 
final states across the parameter space 
allowed by the neutrino mass constraint.
The signal observability at the LHC for both channels  are discussed
in detail in Sec.~IV.
We summarize and conclude in Sec.~V.

\section{Production at hadron Colliders}

\subsection{Drell-Yan production}
As has been already discussed in the literature, the dominant chanel through which
doubly charged scalar pairs can be produced at hadron colliders is the Drell-Yan process
\begin{equation}
q(p_1) \, + \, \bar{q}(p_2) \, \rightarrow \,
H^{++}(k_1)\, + \, H^{--}(k_2),
\end{equation}
In terms of   $y=\hat{\bf p}_1\cdot \hat{\bf k}_1$ in the parton c.m.~frame,  
the  parton level  differential cross section for this process  is
\begin{eqnarray}
\frac{d\sigma}{dy} &=& \frac{16\pi \alpha^2 \beta^3 (1-y^2)}{N_c {s}} 
\Big\{ e_q^2 \,+ \, 
\frac{ {s}}{({s}-M_Z^2)^2+\Gamma_Z^2 M_Z^2}
\, \frac{\cos 2\theta_W}{\sin 2\theta_W}\nonumber\\
&& \times\Big[2 e_q g_V^q  ({s}-M_Z^2)
\, + \, (g_V^{q2}+g_A^{q2}) {s}\  \frac{\cos 2\theta_W}{\sin 2\theta_W}
\Big]
\Big\} ,
\end{eqnarray}
where  $\beta=\sqrt{1-4m_H^2/{s}}$ is the speed of $H^{++}$ in the c.m.~frame.

The QCD corrections to this process have been also computed,
yielding a next-to leading order (NLO) K-factor of the order 
of 1.25 at the LHC energy for the entire mass range between 
200 GeV and 1 TeV \cite{Muhlleitner:2003me}.  


\subsection{Two-photon fusion}
Due to the stronger electromagnetic coupling of the doubly charged scalar,
one may anticipate a sizable two-photon 
contribution to the pair production. Comparing
with a singly charged scalar, the two-photon channel will
have an enhancement factor of 16.
We consider the dominant contribution from the collinear photons and adopt
the effective photon approximation. 
The event rates from effective photon contributions 
can be just added to the Drell-Yan contributions to the  $H^{++} H^{--} X$
final state .

For the process
\begin{equation}
\gamma(p_1)\, +\, \gamma(p_2)\, \rightarrow \,
H^{++}(k_1)\, +\, H^{--}(k_2),
\end{equation} 

\noindent
the matrix element squared is
\begin{equation}
|{\cal M}|^2 = \frac{128\,e^4\,\left( 1 - 2\,{\beta}^2 + 
      {\beta}^4\,\left( 2 - 2\,y^2 + y^4 \right)  \right)
      }{{\left( -1 + {\beta}^2\,y^2 \right) }^2}.
\end{equation}
The corresponding differential cross section can be obtained as
\begin{equation}
\frac{d\sigma}{dy} = \frac{16 \pi \,{\alpha}^2\,\beta \,
    \left( 1 - 2\,{\beta}^2 + 
      {\beta}^4\,\left( 2 - 2\,y^2 + y^4 \right)  \right)
      }{{s}\,{\left( -1 + {\beta}^2\,y^2 \right) }^2}.
\end{equation}
It is interesting to see the different $\beta$-dependence of the cross sections
for the DY and two-photon processes. The DY process undergoes a pure $P$-wave
channel, while the two-photon process contains all $S,\ P,\ D$ waves leading to
a general dependence of $\beta^{2l+1}$ near the threshold.

Furthermore, the two-photon contributions may arise from 
both elastic and inelastic processes, including the 
semi-elastic case where it is elastic
on one side and inelastic on the other. 
The total cross-section can thus be written
as \cite{Drees:1994zx}

\begin{equation}
\sigma_{\gamma\gamma} = 
\sigma_{\rm elastic} + \sigma_{\rm inelastic} +\sigma_{\rm semi-elastic}
\end{equation}

\noindent
where the elastic channel is the photon radiation off a proton, and the inelastic channel
is that off a quark parton. The semi-elastic is the product of both. They
are given by 
\begin{eqnarray}
\sigma_{\rm elastic} &=& \int^1_\tau dz_1 \int^1_{\tau/z_1} dz_2  f_{\gamma/p}(z_1) f_{\gamma/p'}(z_2)\sigma(\gamma \gamma \rightarrow H^{++} H^{--}), \qquad \tau={4m^2\over S},  \\
\sigma_{\rm inelastic} &=& \int^1_\tau dx_1 \int^1_{\tau/x_1} dx_2 \int^1_{\tau/x_1/x_2} dz_1 
\int^1_{\tau/x_1/x_2/z_1} \\
&& dz_2 f_q(x_1) f_q'(x_2) f_{\gamma/q}(z_1) 
f_{\gamma/q'}(z_2)\sigma(\gamma \gamma \rightarrow H^{++} H^{--})\\ 
\sigma_{\rm semi-elastic} &=& 
\int^1_{\tau} dx_1 \int^1_{\tau/x_1} d z_1\int^1_{\tau/x_1/z_1} dz_2 f_q(x_1) f_{\gamma/q}(z_1) f_{\gamma/p'}(z_2)\sigma (\gamma \gamma \rightarrow H^{++} H^{--}).~~~
\end{eqnarray}
We employ the frame structure functions following reference \cite{Drees:1994zx}:
\begin{eqnarray}
&& f_{\gamma/q}(z) = {\alpha_{\rm em}\over 2 \pi}{{1+(1-z)^2}\over z} {\rm ln}(Q^2_1/Q^2_2)\\
&& f_{\gamma/p}(z) = {\alpha_{\rm em}\over 2 \pi z}(1+(1-z)^2) \left[{\rm ln}A -{11\over 6}+{3\over A}-
{3\over 2A^2}+{1\over 3A^3}\right] \\
&& A = 1 + {0.71~{\rm GeV}^2\over Q^2_{\rm min}} \\
&& Q^2_{\rm min} =  -2m^2_p +{1 \over 2s}\left[ (s+m^2_p)(s-zs+m^2_p)-(s-m^2_p)
\sqrt{(s-zs-m^2_p)^2-4m^2_pzs}\ \right].~~~~~
\end{eqnarray}

The treatment in this section has ignored  the channels from other gauge boson
fusions such as $W^*,\ Z^*,\ \gamma^*$ (virtual in general). 
Our estimates reveal that the production through $W$-fusion channel is
rather suppressed and more so for $Z$-fusion.
The cost of using the effective photon approximation rather than 
calculating the full $2\to 4$ subprocess is the loss 
of the potential tagging jets in the forward-backward
regions. Fortunately, due to the rather clean charged leptonic 
final states, we do not require such  jet tagging to identify the signal. 

\subsection{Numerical results}

\begin{figure}[t]
\includegraphics[scale=1,width=8.15cm]{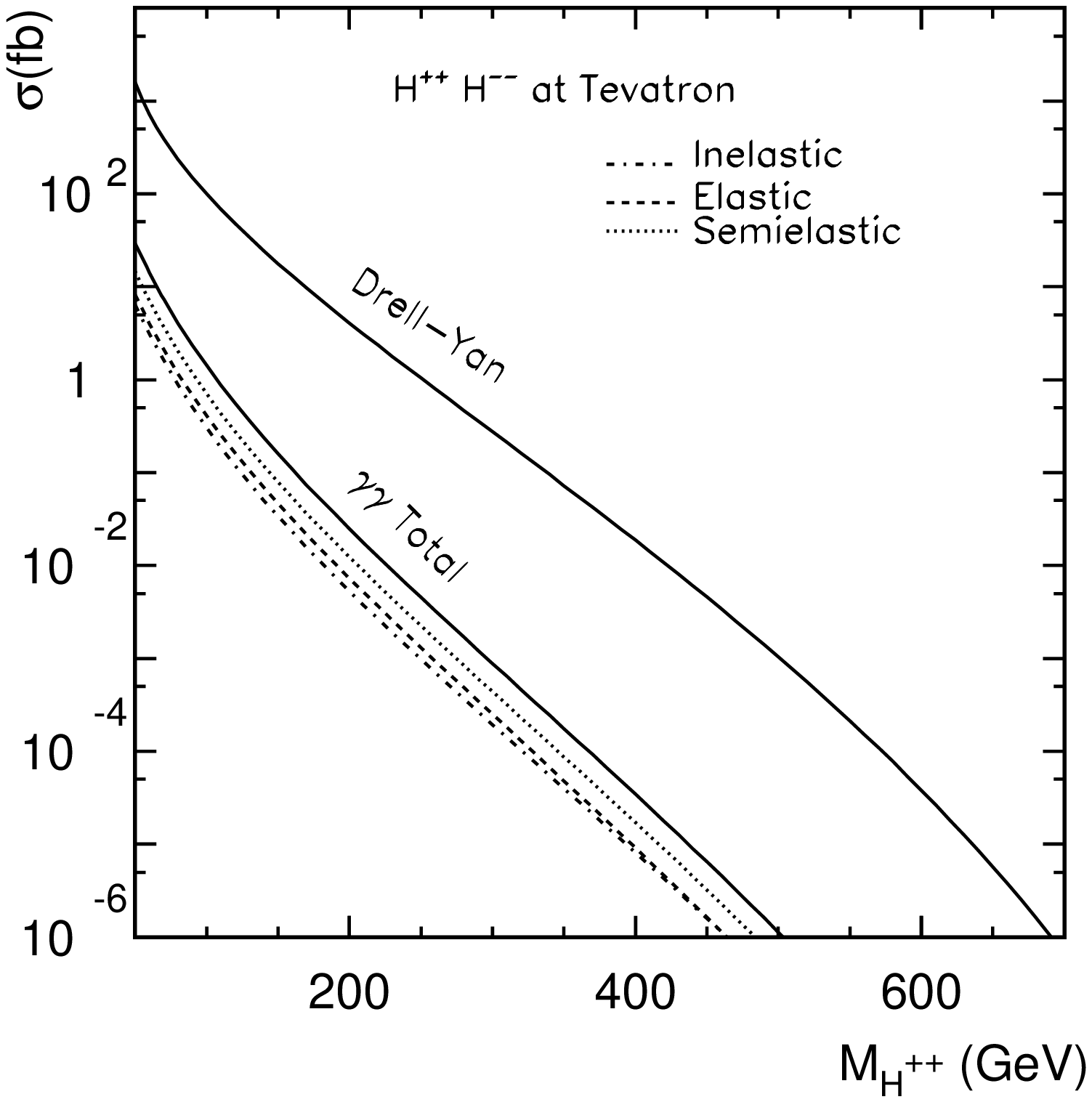}
\includegraphics[scale=1,width=8.15cm]{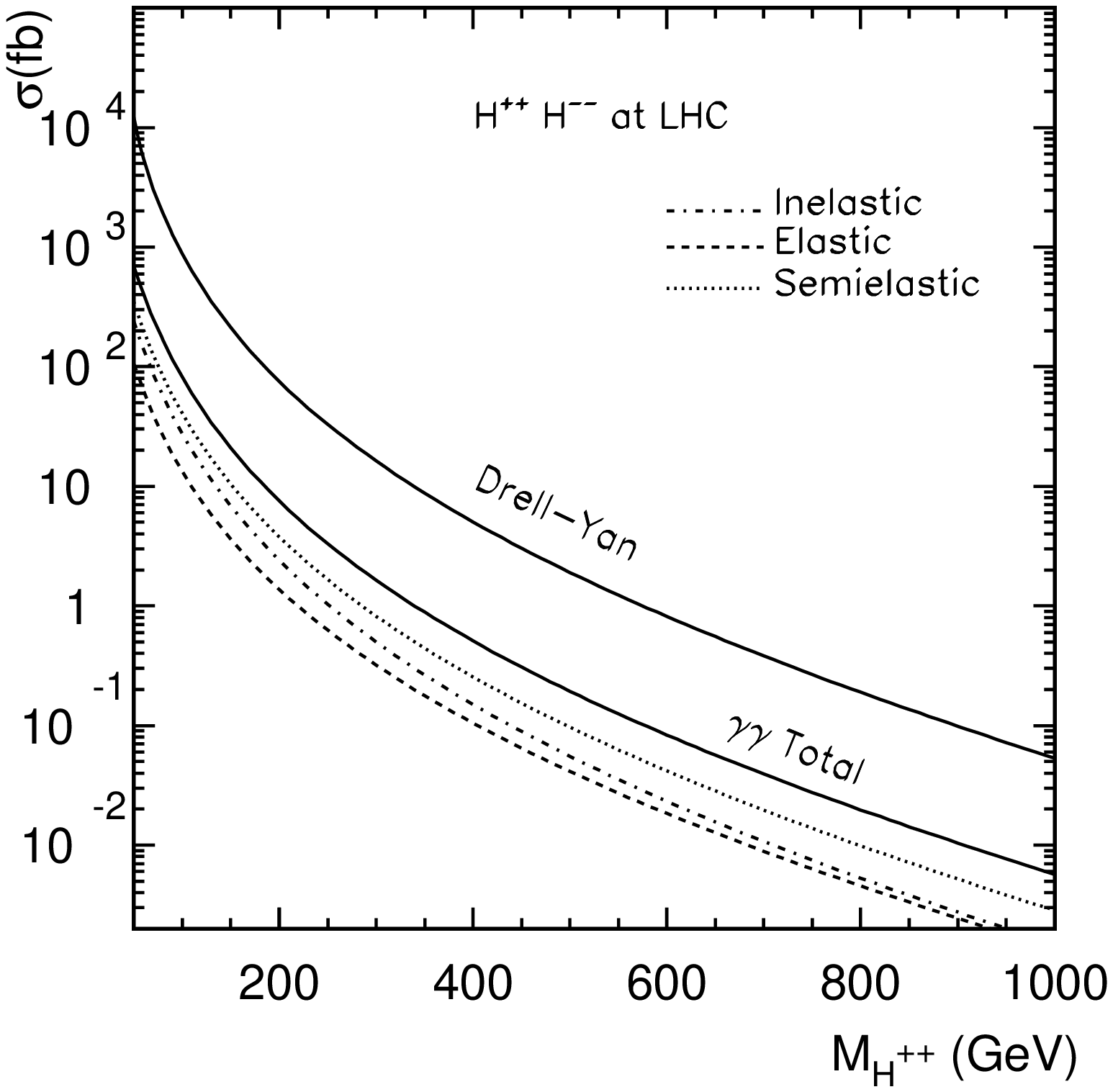}
\caption{Production rates for the doubly charged Higgs pair at the  Tevatron 
(left) and the LHC (right), in the leading order Drell-Yan and two-photon 
(semi-elastic, elastic, and inelastic) channels.} 
\label{rate}
\end{figure}

\begin{figure}[t]
\includegraphics[scale=1,width=8.5cm]{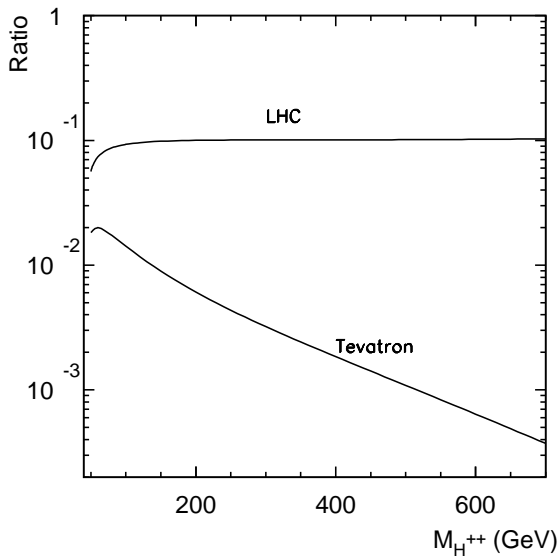}
\caption{The ratio between $\sigma_{\gamma\gamma}$ and leading order
$\sigma_{\rm DY}$ at the LHC and the Tevatron.}
\label{ratio}
\end{figure}

In Fig.~\ref{rate} the total cross-sections are plotted against the mass of the doubly
charged scalar,  showing the Drell-Yan as well as different types of the two-photon
processes. 
The highest lower  bound on a doubly charged Higgs mass 
from the Tevatron search with an integrated luminosity of
240 pb$^{-1}$ is about 136  GeV (though in the specific context of
a left-right symmetric model). 
It can be seen from Fig.~\ref{rate} that with 10 fb$^{-1}$ at the Tevatron, 
the mass reach may be extended approximately to 250 GeV.
At the LHC, the production rate is increased by a factor of 20 over that at the
Tevatron, reaching the order of femtobarns for a Higgs mass upto as much 
as 600 GeV, and of 0.1 fb for a mass of 1 TeV.
We have also shown the three classes of contributions from the two-photon channel separately.
It is interesting to note that the rate of the elastic process is larger than that of the inelastic
at the Tevatron energies, while the pattern 
is reversed at the LHC energies. This is because the probability of
a proton remaining unbroken after the emission of a photon in an elastic process 
is smaller at the LHC. The semi-elastic contribution is numerically the highest in  both 
cases,  because of a factor of 2 for the initial state interchanges.
The ratio of the two-photon contribution relative to the Drell-Yan channel 
is shown in Fig.~\ref{ratio}. In general, the two-photon contributions at LHC 
remain about 10\% of that of the Drell-Yan process, while the fraction is much smaller
for the Tevatron, the reason being that is the photon needs to come with a larger
momentum fraction of the parent parton in the latter case, for which the distribution
function is relatively suppressed.
It should be noted that, while the Drell-Yan cross-section has a 
next-to-leading order (NLO) QCD K-factor of about 1.25 \cite{Muhlleitner:2003me}, 
only the leading order cross-section 
has been presented in these figures. This is for comparison of the correction
to this cross-section due to the two-photon process and the strong correction
mentioned above.  Higher order corrections to the two photon 
electromagnetic process are rather small. 
CTEQ6L parton distribution functions have been used in the calculation, and 
the factorization scale is set at half the subprocess center-of-mass energy 
$\sqrt{{s}}/2$. 

\begin{figure}[t]
\includegraphics[scale=1,width=8.15cm]{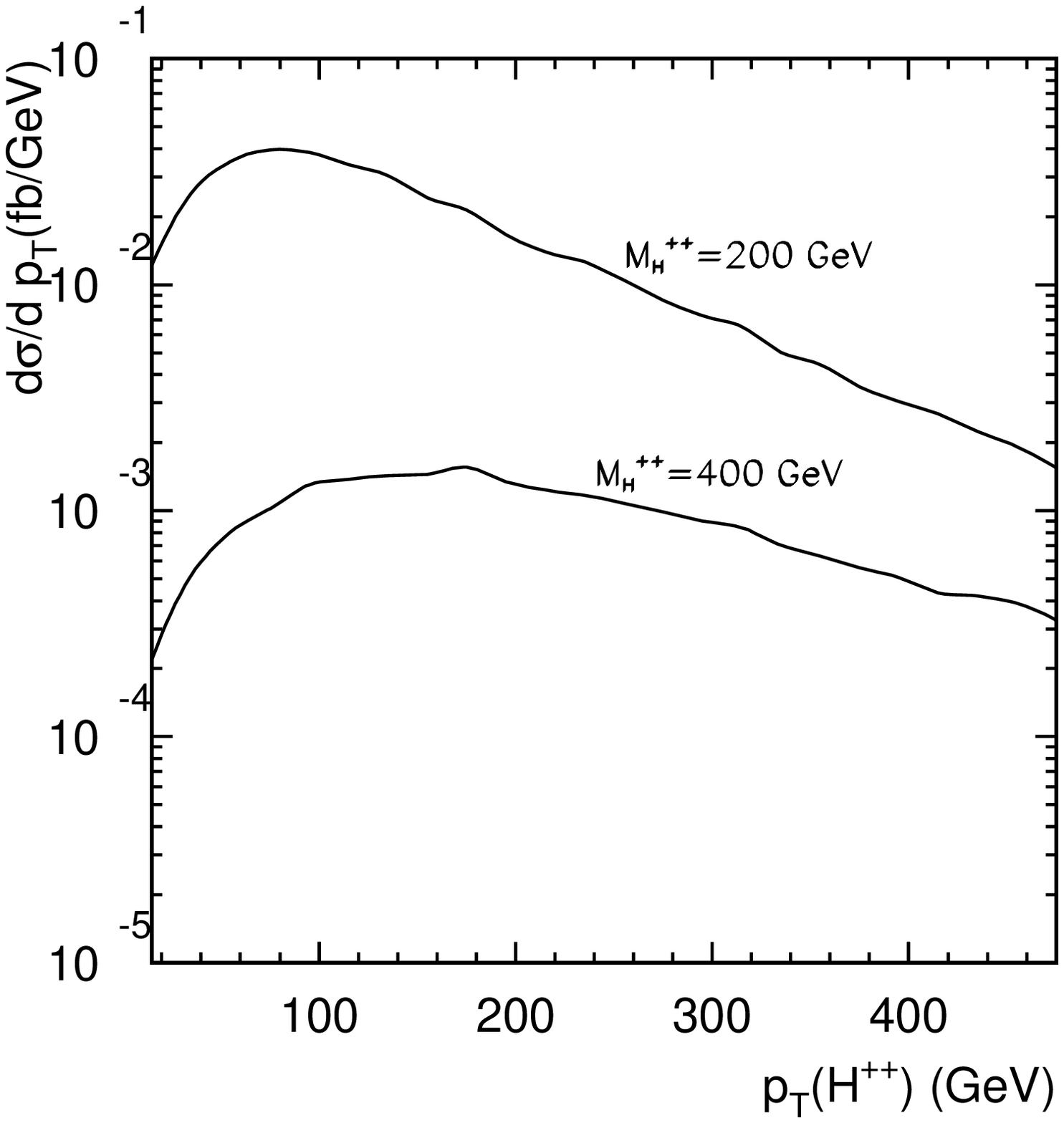}
\includegraphics[scale=1,width=8.15cm]{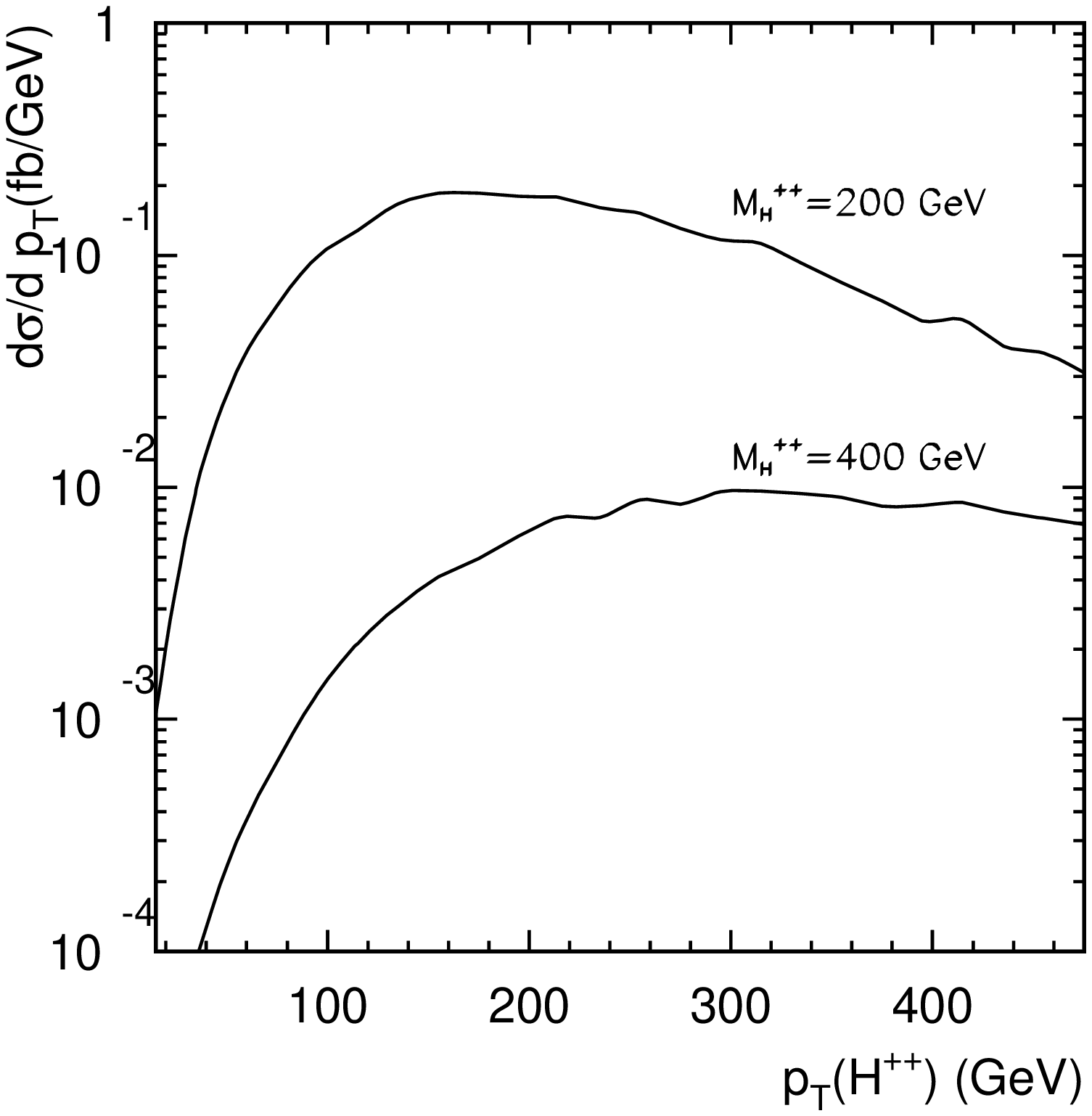}
\caption{Transverse momentum distributions of the doubly charged Higgs produced 
in the two-photon (left) and Drell-Yan (right)  channels at the LHC.}
\label{pt}
\end{figure}

\begin{figure}[t]
\includegraphics[scale=1,width=8.15cm]{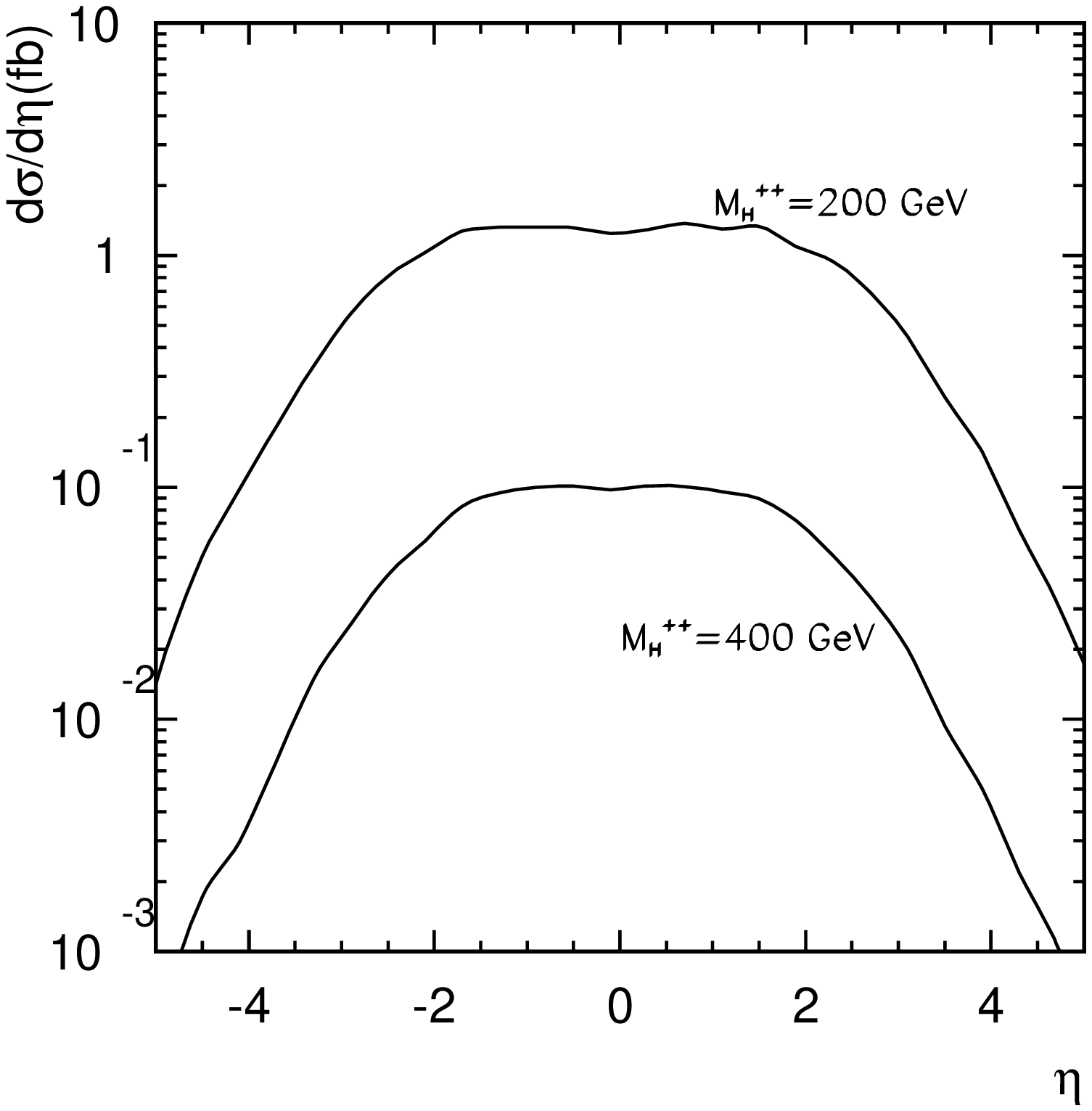}
\includegraphics[scale=1,width=8.15cm]{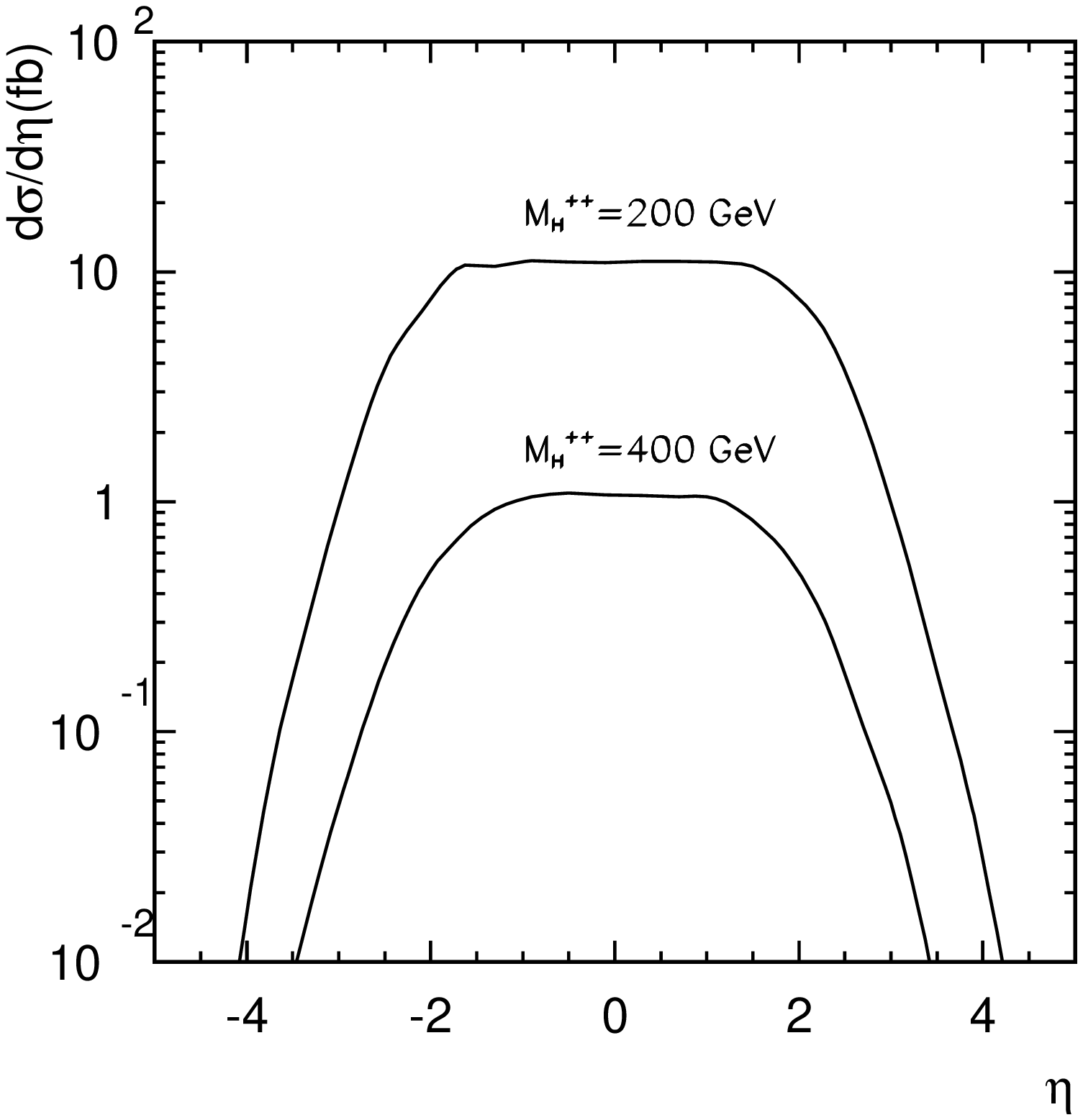}
\caption{Rapidity distributions of the doubly charged Higgs produced in the two-photon (left)
and Drell-Yan (right)  channels at the LHC.} 
\label{eta}
\end{figure}

%
We now examine some kinematic features of the
two processes considered above.
As shown in Fig.~\ref{pt},  
while the transverse momentum ($p_T$) of the scalars
tends to peak at values that are  significant fractions of its mass, 
Drell-Yan production leads to scalars with harder $p_T$ 
distributions than those from the two-photon fusion. This is 
due to the forward-backward
nature of the $t,\ u$-channels in the two-photon process at high energies.
The rapidity $(\eta)$ distributions in Fig.~\ref{eta}
reflect central peaking for both channels, and the distributions
are broader for the two-photon process. On the whole,
in spite of some qualitative differences between these two channels, 
it is not possible effectively to separate the two production mechanisms through the
kinematical considerations. In our subsequent analysis, we will
add them up incoherently. 


\section{Decays of the doubly charged Higgs}

The doubly charged Higgs may in principle decay into all of the following 
two-body final states:
\begin{eqnarray}
&& H^{++} \to \ell_i^+ \ell_j^+ ,\quad H^{++} \to W^+ W^+ . \nonumber\\
&& H^{++} \to H^+ W^+, \quad H^{++} \to  H^+ H^+ . \nonumber
\end{eqnarray}
The third and fourth channels depend on the mass splitting among the members of
the triplet, and also the scalar potential. The consequence of the third channel being
kinematically allowed (which makes it the dominant decay, since it is driven by
the SU(2)$_L$ gauge coupling)  has been discussed, for example, in reference  
\cite{Chakrabarti:1998qy},  in the context
of a linear collider.  Here we take a conservative approach and 
assume that the mass splitting within the triplet is very small 
as is the case, for example, in Little Higgs models.  Thus only the first two
decays listed above occur in our scenario,\footnote{
The omission of the $H^\pm W^\pm$ mode will affect neither our analyses nor the
results for the individual channels $\ell^\pm \ell^\pm$ and $W^\pm W^\pm$, 
since the branching ratios in these channels have been used by us
as free parameters. However, the inclusion of this mode can in principle 
alter the overall discovery limit, 
depending upon the mass parameters.}  
the corresponding widths being given by \cite{Han:2005nk}

\begin{eqnarray}
	\Gamma (H^{++}\rightarrow \ell^+_i \ell^+_j) &=& 
	\frac{1}{4\pi (1+\delta_{ij})}|\yll^{ij}|^2 M_{H^{++}} \ ,
		 \nonumber \\
	\Gamma (H^{++}\rightarrow W^+_T W^+_T) &=& \frac{1}{4\pi }
	\frac{g^4v^{\prime 2}}{M_{H^{++}} }
	\frac{\lambda^{\frac{1}{2}}(1 ,r^2_W ,r^2_W)}{\sqrt{4r_W^2 + 
	\lambda (1 ,r^2_W ,r^2_W)}} \approx  
	\frac{g^4v^{\prime 2}}{4\pi M_{H^{++}} },
		\nonumber \\
	\Gamma (H^{++}\rightarrow W^+_L W^+_L) &=& \frac{1}{4\pi }
	\frac{g^4v^{\prime 2}}{2M_{H^{++}} }
	\frac{\lambda^{\frac{1}{2}}(1 ,r^2_W ,r^2_W)}{\sqrt{4r_W^2 + 
	\lambda (1 ,r^2_W ,r^2_W)}} \frac{(1 - 4r_W^2)^2}{4r_W^4}
	\approx  \frac{v^{\prime 2} M_{H^{++}}^3}{2 \pi v^4}.
\end{eqnarray}

\noindent
in terms of the kinematic function 
$\lambda (x,y,z) = x^2 + y^2 + z^2 - 2xy - 2xz - 2yz$ and the scaled
mass variable $r_W = M_W/M_{H^{++}}$. The approximate forms are
valid for $M_{H^{++}} \gg M_W$.  
The subscripts $T$ and $L$ denote the transverse and longitudinal 
polarizations of the $W$ boson.  
The longitudinal $W$ final state becomes dominant at higher $M_{H^{++}}$.

\begin{figure}[t]
\includegraphics[scale=1,width=8.15cm]{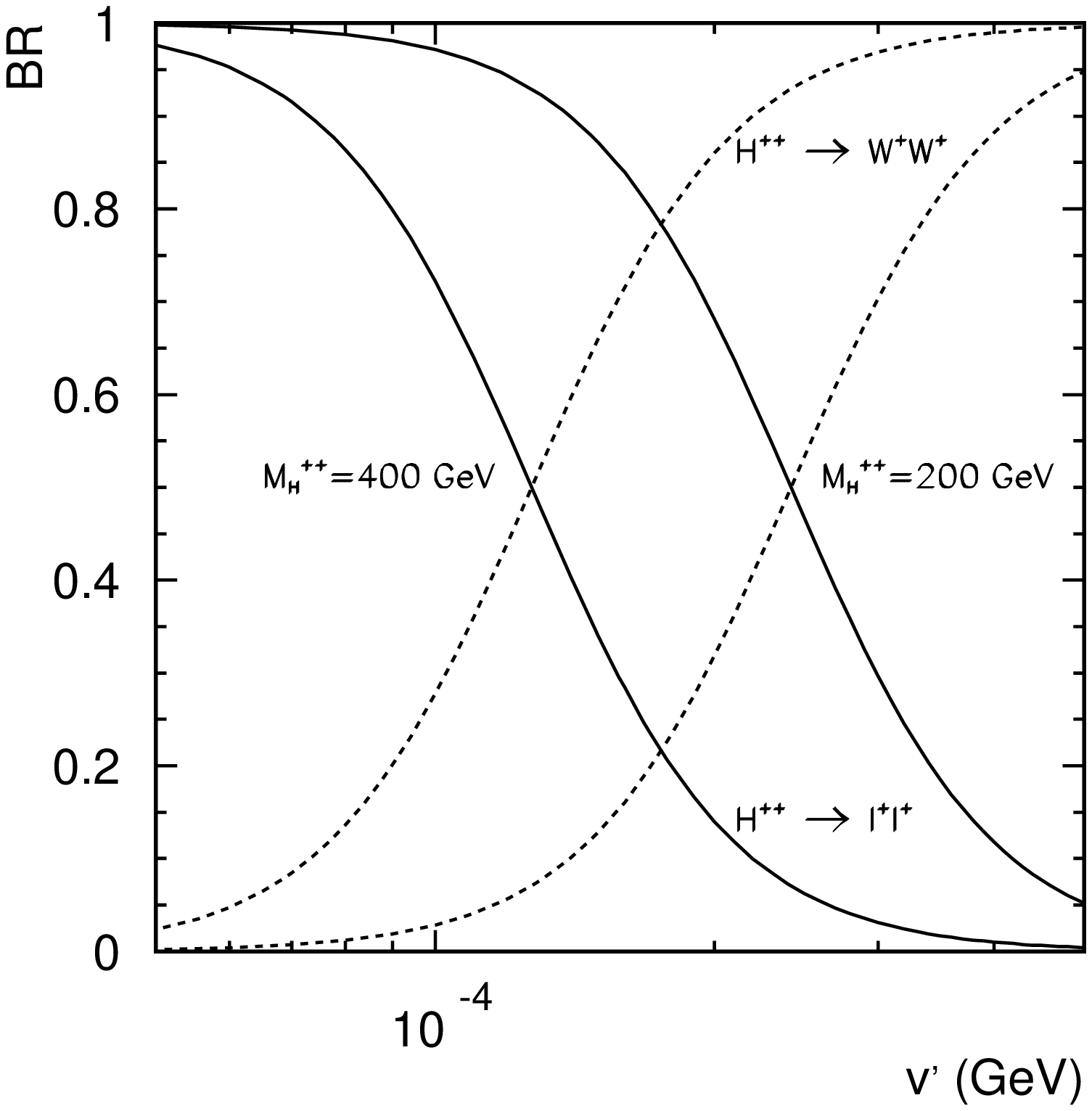}
\includegraphics[scale=1,width=8.15cm]{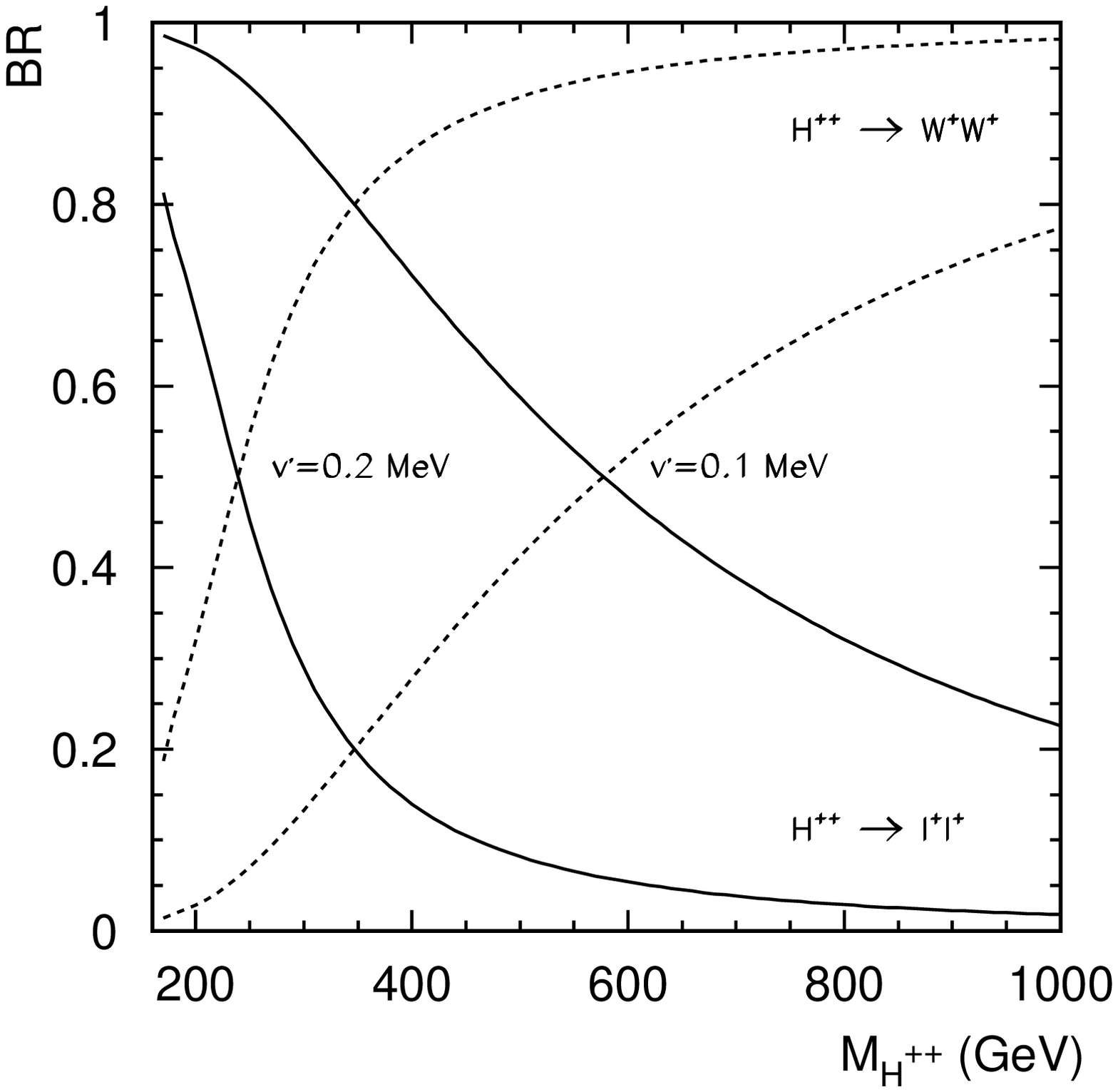}
\caption{Branching fractions for  $H^{++}\to \llpp\ (e^+e^++\mu^+\mu^++\tau^+\tau^+)$  and
$H^{++}\to W^{+}W^{+}$ as functions of (left) the triplet vev, for
two values of the doubly charged Higgs mass, and (right) the Higgs
mass for two choices of $v'$. }
\label{br}
\end{figure}

The relative strengths of the two types of decays ($\ell^{+}\ell^{+}, W^{+}W^{+}$)
depend on the couplings $Y_{\ell\ell}$ as well the triplet vev $v'$, which 
has to be less than 1 GeV in order to prevent large tree-level 
contributions to the $\rho$-parameter unless additional model assumptions are made.  
While  treating $Y_{\ell\ell}$ and $v'$
as completely free parameters can lead to practically any relative 
strength between
the two sets of final states, we have been guided by 
the additional constraint of 
neutrino mass generation, to saturate the bound of Eq.~(\ref{yvev}). 
%
%
Of course, ${\yll}v'$ can be even smaller if there are right-handed
neutrinos in addition, for instance, a Type II 
seesaw mechanism is operative. 
We shall further comment on this possibility at the end of this section. 
In Fig.~\ref{br} we show the branching fractions 
for the $\llpp\  (e^+e^++\mu^+\mu^++\tau^+\tau^+)$  and   $W^{+}W^{+}$ 
decay modes as functions of the triplet vev (left)
and  the Higgs mass (right), keeping the overall constraint from
neutrino mass mentioned above.
With a higher mass of $H^{++}$,  the $W^{+}W^{+}$ mode  overtakes  
$\llpp$  sooner due to the fast growing $W_LW_L$ mode, even for 
a relatively smaller value of  $v' $.  
While the explanation is obvious from the expressions of the decay widths, this 
underlines  the importance of exploring the latter 
mode at LHC, in addition to the easily identifiable 
like-sign lepton pair signal.

\begin{figure}[t]
\includegraphics[scale=1,width=8.5cm]{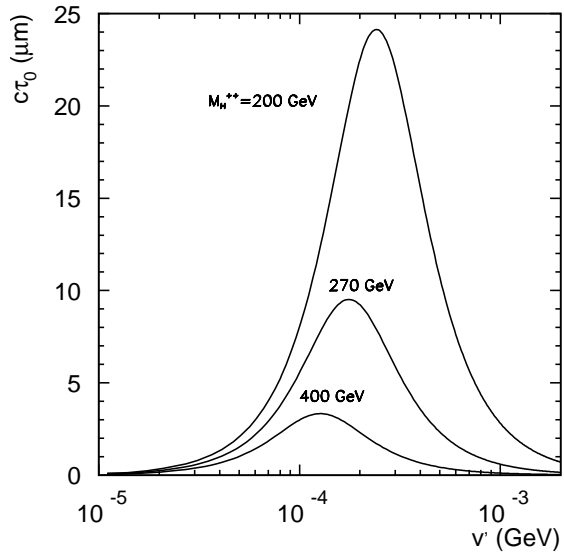}
\caption{The proper decay length $c\tau_0$ in units of $\mu$m for the doubly 
charged Higgs as a function of the triplet vev, within the neutrino mass constraint. }
\label{dl}
\end{figure}

Since $\yll$ and $v'$ are both quite small, it is natural to ask whether the decay
width of $H^{++}$ can be so small that it becomes 
quasi-stable in the collider experiments.  As has been mentioned earlier, 
a long-lived doubly charged Higgs has been looked for at the Tevatron, in terms
of highly ionizing tracks and muon-like penetration beyond the electromagnetic 
and hadron calorimeters \cite{Acosta:2005np}.  In Fig.~\ref{dl}, 
we plot the the proper decay length $c\tau_0$ with $\tau_0$ 
being the proper lifetime of $H^{++}$,  keeping within the constraint on
$Y_{\ell\ell} v'$ from neutrino mass to saturate Eq.~(\ref{yvev}). 
It can be seen that, for a light scalar $M_{H^{++}}^{}<270$ GeV with  certain values of $v'$, 
the proper decay length may exceed 10 microns. Such a decay length can be  enhanced 
to a visible displaced secondary vertex by the appropriate boost $\beta \gamma=p/M$. 
The length goes further down for a more
massive scalar. Thus it can be concluded that a 
doubly charged Higgs cannot be long-lived, or quasi-stable, 
on the scale of a collider detector if its $\Delta L = 2$ 
coupling has to be the mechanism
operative for the generation of neutrino masses. 
The observation of a long-lived $H^{++}$
should therefore tell us that something over and above 
the $\Delta L = 2$ interaction
is active for generating neutrino masses.

It should  also  be remembered that $\yll^{ij}$ can have six 
independent elements
(assuming a real symmetric Majorana neutrino mass matrix) in general. 
Values of these elements are highly model-dependent and they can lead to a 
 wide number of possibilities in flavor diagonal as well as 
off-diagonal decays
of the $H^{++}$.   While there is very little guideline as to 
the choice of these values,
it should be emphasized that  actual observation of the decays 
into different flavor
combinations can actually give us information about 
the structure of the neutrino mass matrix. 
In order to convey the general features of the collider analysis, 
we assume a diagonal structure of ${\yll}^{ij}$, with the same value for 
all three flavors, as shown in Fig.~\ref{br}.
We include only the electronic and muonic 
branching ratios added together while
predicting the signal events at the hadron colliders.  This analysis can 
always be translated into one with
a variety of structures for 
the $\yll^{ij}$, without any serious difference in
the results presented here.

\section{Prediction of events at the LHC}

\begin{figure}[t]
\includegraphics[scale=1,width=8.5cm]{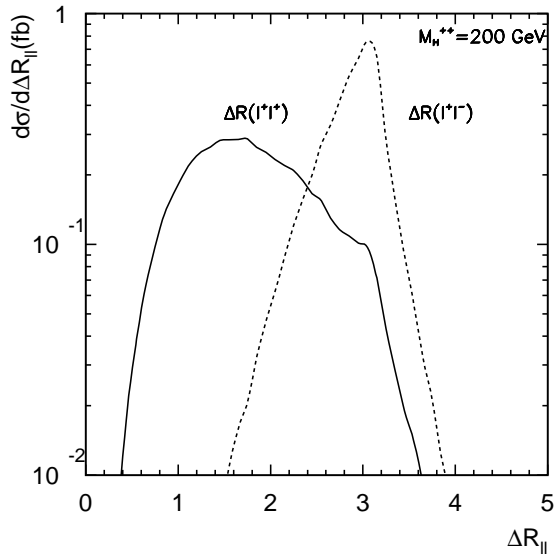}
\caption{$\Delta R_{\ell\ell}$ distributions for same-sign  (solid)
opposite-sign (dotted) dileptons
in the four-lepton final state. For opposite-sign dileptons, the isolation between the
two harder leptons of each sign has been selected.}\label{dr}
\end{figure}

\subsection{Like-sign dilepton pairs}

We first consider the most spectacular final state, namely, 
\begin{equation}
pp \to H^{++} H^{--} \to \llpp\ \llmm\ (\ell=e,\ \mu),
\end{equation}
driven
by the decay of the $H^{++}$ into lepton pairs. As mentioned earlier, 
we have assumed
the same branching ratio for flavor diagonal 
decays into the three lepton channels, and included
the electron and muon pairs as  our characteristic signals. 
The branching ratio into  lepton
pairs has been taken as a free parameter, since,  
given   the neutrino mass constraint, the
decay rate into $W^{+} W^{+}$  is automatically determined by it.  

We consider both electrons and muons. To simulate the detector effects on
the energy-momentum measurements, we smear the electromagnetic energy
and the muon momentum tracking by a Gaussian distribution whose width is \cite{ATLAS}
\begin{eqnarray}
{ \Delta E\over E} &=& {a_{cal} \over \sqrt{E/{\rm GeV}} } \oplus b_{cal}, \quad 
a_{cal}=10\%,\  b_{cal}=0.4\% ,
\label{ecal}\\
{\Delta p_T\over p_T} &=& {a_{track}\ p_T \over {\rm TeV}} \oplus {b_{track}\over \sqrt{\sin{\theta}} }, \quad
 a_{track}= 36\%,\ b_{track} =1.3\%.
\end{eqnarray}

The signal consists of two like sign dilepton pairs with 
the same invariant mass. As can be 
seen from Fig.~\ref{dr}, the $\Delta R$ 
distribution (where $\Delta R^2 = \Delta \eta^2  + \Delta \phi^2$)
of the like-sign pairs peaks at a smaller value than that 
of opposite sign dileptons, thereby
indicating a spatial separation of the pairs, 
at least when the doubly charged scalars are
sufficiently boosted.  We have checked that this 
separation remains noticeable well beyond $M_{H^{++}} \simeq 400 - 500$ GeV. 
In addition, we have employed the 
following acceptance criteria for the events:
\begin{itemize}
\item A veto on any opposite sign dilepton pair invariant mass being close to the
$Z$-boson mass: 
$| m({\ell^+ \ell^-}) - M_Z |>15~{\rm GeV}$.
 \item $p_T(\ell) > $ 15 GeV, and the hardest leptons has 
 $p_T(\ell_{hard}) > $ 30 GeV.
\item $|\eta(\ell)| < $ 2.8.
\end{itemize}

The above signal has {\it prima facie} no standard model background. However,
fake backgrounds may arise from $W$ bosons in conjunction with
misidentified jets and/or leptons from heavy flavor decays.  In order to 
suppress such backgrounds, we recommend the following steps:

\begin{itemize}
\item Reconstruct the invariant mass of each like-sign dilepton pair,
and insist on their being equal within about 5 per cent. While the smearing 
procedure is seen to keep the invariant masses of the signal pairs equal
well within this limit,  backgrounds are largely eliminated by it.

\item Demand a minimum $\Delta R$-isolation of 0.5 between each lepton and 
any nearby jet.

\item Demand little missing transverse energy:
$\cancel{E}_T <25$ GeV.
\end{itemize}

\begin{figure}[t]
\includegraphics[scale=1,width=8.5cm]{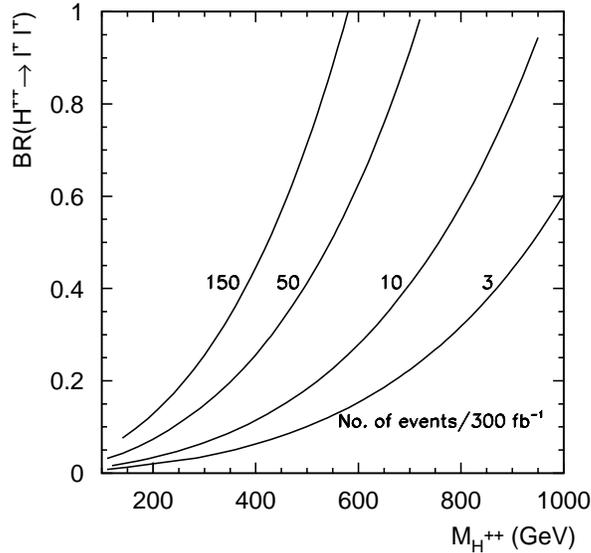}
\caption{Equal event contours in the 
${\rm BR}(H^{++}\to \ell^{+}\ell^{+})-M_H^{++}$ plane,
 including the cuts discussed in the text,  
 for an integrated luminosity of  300 fb$^{-1}$.}
 \label{evt}
\end{figure}

In Fig.~\ref{evt}   we present event contours in the space spanned 
by the $H^{++}$ mass and the branching fraction $BR(H^{++}\to \llpp )$ 
for an integrated luminosity of 300 fb$^{-1}$.
The Drell-Yan contribution has been  multiplied here by the NLO K-factor 1.25, and 
the two-photon contribution has also been included. 
In the absence of backgrounds after all of the cuts discussed above,  
the contour corresponding to three events 
can be taken as the $99\%$ C.L. discovery limit 
for an integrated luminosity of  300 fb$^{-1}$. 
The event rates suggest a high degree of detectability of the signal, 
even for rather small branching ratios.  Given the acceptance criteria listed above, 
it can be  concluded from the contours that a doubly charged Higgs 
up to the mass range of 1 TeV can be detected at the LHC if the $\llpp$  
decay mode has a branching ratio  on the order of $60$ per cent.
On the lower side, even a branching ratio
of 10 (5) per cent allows a $99\%$ C.L. search limit of 500 (350) GeV. 
A recent similar study has appeared \cite{recentLL} for the DY process and the
leptonic decay mode of $H^{++}$ only. Their numerical results, wherever overlapping, 
are in agreement with ours.

\begin{figure}[t]
\includegraphics[scale=1,width=8.15cm]{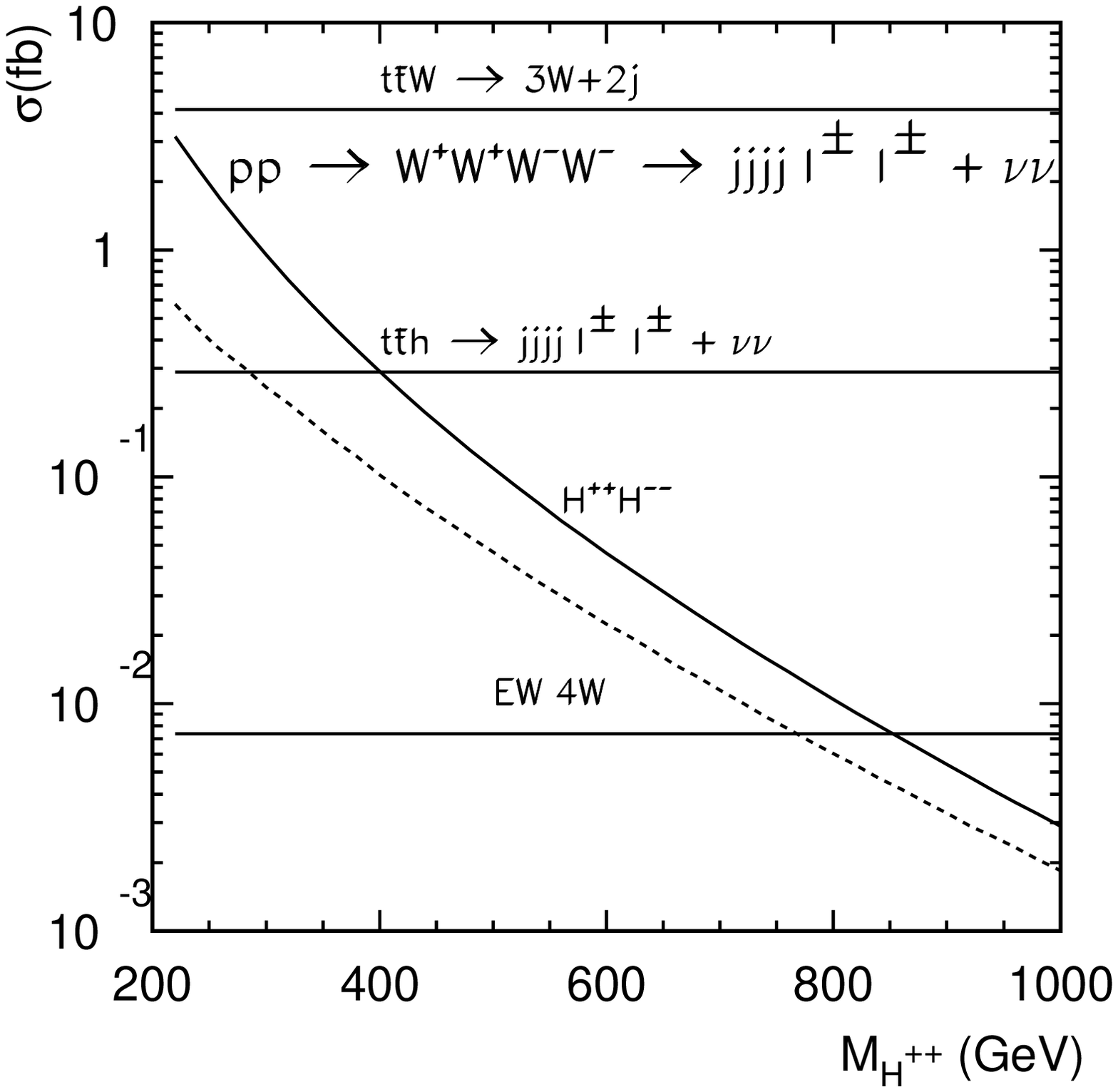}
\includegraphics[scale=1,width=8.15cm]{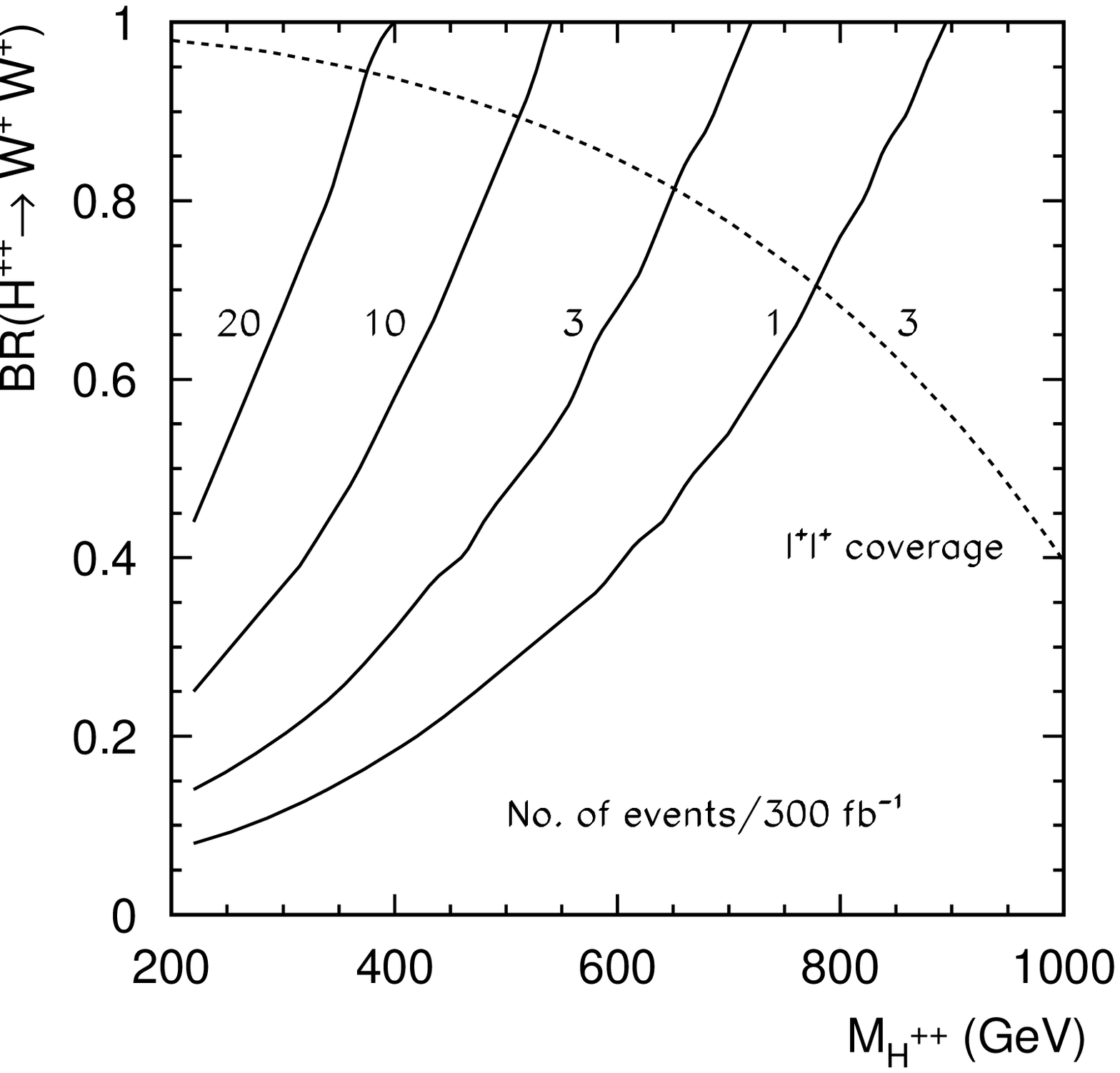}
\caption{(a) Left: Total cross sections for $jjjj\ell^\pm \ell^\pm + \cancel{E}_T$ events.
 The solid (dotted) line  corresponds to event rates 
 coming from each doubly charged scalar decaying into two like-sign $W$'s
 without (with) basic cuts as 
described in the text. Event rates from various types of background are also shown, without the basic cuts. (b) Right: Equal event contours in the ${\rm BR}(H^{++}\to W^{+}W^{+})-M_H^{++}$ plane,
including the cuts discussed in the text,  
 for an integrated luminosity of  300 fb$^{-1}$. The dashed curve is a reproduction of
 the 3-event contour in Fig.~\ref{evt}. }
 \label{ww}
\end{figure}

\subsection{Like-sign $W$-pairs}

Next, let us consider the $W^{+}W^{+}$ channel. 
As can be seen from Fig.~\ref{br}, 
this channel may become dominant even for 
rather small values of the triplet vev,
especially if the $H^{++}$ mass is on the higher side.   
It is therefore desirable to devise search strategies for this channel, 
leading to a final state consisting of $W^{+}W^{+} W^{-}W^{-} $.  
In order to confirm the nature of a doubly charged state, and
to reveal its resonant production, we propose to reconstruct the events
by looking for two like-sign $W$'s through a pair of like-sign dileptons and 
the remaining two in their hadronic decays which allow complete reconstruction
of $M_{H^{++}}$.
%
The branching fraction for this decay chain is taken to be
\begin{equation}
{\rm BR}(W^{+}W^{+} W^{-}W^{-}\to  \ell^\pm_i\nu\ \ell^\pm_j \nu\ 4j ) \approx 2\ ({2\over 9})^2\ ({6\over 9})^2
\approx 4.4\% .
\label{br4w}
\end{equation}
The predicted cross sections including the above branching fraction 
are shown in Fig.~\ref{ww}(a),  assuming  BR($H^{++} \to W^+ W^+$) to be 100\%. 
The rather small branching fraction in Eq.~(\ref{br4w})  is a price we pay
for cleaner signals, and consequently the event rate tends to 
be low for a high $H^{++}$ mass.
For instance, it may only have a handful events for $M_{H^{++}} \sim 1$ TeV 
 with an integrated luminosity of 300 fb$^{-1}$, even before any acceptance cuts.

We again start with some ``basic cuts", where 
the leptons are subject to the same acceptance criteria as before.
In addition, we demand 
\begin{itemize}
\item $\cancel{E}_T > 40\ {\rm GeV}$
\item $p_T(j) \geq  30\ {\rm GeV}$, $|\eta(j)| \leq 3.0.$ 
\end{itemize}
\noindent 
The jet energies are also smeared using the same Gaussian formula \cite{ATLAS}
as in Eq.~(\ref{ecal}), but with 
\begin{equation}
a=80\%,\quad  b=15\%.
\end{equation}
As seen from the solid and dotted curves in Fig.~\ref{ww}(a),  
these criteria do not affect the signal in a significant way, especially for heavier masses.
In Fig.~\ref{ww}(b) we translate the signal cross section to 
equal event contours in the ${\rm BR}(H^{++}\to W^{+}W^{+})-M_H^{++}$ plane,
including the cuts discussed above and both $\llpp$ and $\llmm$, 
for an integrated luminosity of  300 fb$^{-1}$. 
For comparison, we reproduce  the 3-event contour in Fig.~\ref{evt} by the darshed
curve, assuming ${\rm BR}(H^{++}\to \ell^+\ell^+)=1 - {\rm BR}(H^{++}\to W^{+}W^{+})$.
This illustrates the complementarity between these two channels for different values
of the BR's.

As for the SM backgrounds, the purely electroweak 
production of $4W$ final states is rather modest \cite{Barger:1989cp}, 
as indicated in Fig.~\ref{ww}(a) with the branching fraction 
of Eq.~(\ref{br4w}) included.
A next background leading to $4W$ final states from 
$t\bar t\ t\bar t$ production \cite{Barger:1991jn} is larger but 
the kinematics will look very 
different from the signal processes due to the
four extra hard $b$ jets. It is easy to demonstrate that these backgrounds
can be effectively suppressed well below the signal by judicial kinematical cuts.
%

The largest background arises from $t\bar{t}W$ production. 
In addition,  if the $WW^{(*)}$ mode from the SM Higgs boson 
($h$)  has a substantial branching ratio, then the $t{\bar{t}} h$ 
production \cite{Barger:1991jn,dawson}
may constitute a severe background. 
As shown in \cite{dawson}, this production cross-section for $t{\bar{t}}h$
can be of the order of 150 fb for a Higgs mass of 200 GeV when it has
unsuppressed decays into two $W'$s, and this can contribute enough
4-jet plus like-sign dilepton events to swamp our signals. 
We have found that the event selection strategies which can suppress the
leading $t\bar{t}W$ background are also effective in reducing the
fake events coming from $t{\bar{t}}h$. Therefore,
we outline below a step-by-step 
procedure for reducing the leading background, as summarised in
Table \ref{Tab:I}.  The relatively smaller backgrounds of $4t$ and $4W$ are 
also taken care of by the same procedure. 
For illustration, we first choose a doubly-charged 
Higgs mass of  300 GeV. In the results presented, the branching ratio 
for $H^{++}\to W^+ W^+$  has been taken as unity. 

First of all, we wish to select events with four jets and thus we 
veto events with additional central jets in the region, 
$$|\eta^{veto}(j)| <3,\ p_T^{veto}(j)>30~{\rm GeV}.$$ Then we
experiment with progressively stronger cuts on the (leading) lepton
and jet $p_T$. Furthermore, the four jets will pair 
up to reconstruct two $M_W$ peaks, and we thus demand 
$|M_{j_{1}j_{2}} - M_W| <15$ GeV for the pair closest to the 
$M_W$ peak and the other pair at the same time. 
As seen from the table, this  reduces both the $t\bar{t}W$ and 
$t\bar{t}h$ backgrounds rather drastically, but not to the extent
necessary for discerning the signal. Therefore, one should
note that the pair production of the heavy Higgs bosons has a high energy
threshold. We accordingly define a cluster transverse mass for the whole
system
\begin{equation}
M_{cluster}  =  \sqrt{m^2_{4j} + (\sum {\vec{p_T}^j})^2}  +
 \sqrt{m^2_{\ell\ell} + (\sum{\vec{p_T}^\ell})^2}  + {\cancel{E}}_T
\end{equation}
which should start from a threshold around $2M_{H^{++}}$.
We  impose a cut of $600$ GeV on this variable.
For the $H^{\pm\pm}$ kinematics on the leptonic side, we can only define a 
transverse mass due to the missing neutrinos
\begin{equation}
M_T  = \sqrt{ (\sqrt{m^2_{\ell\ell}  + (\sum {\vec{p_T}^\ell})^2} + \cancel{E}_T)^2  - 
  (\sum{\vec{p_T}^\ell} + \vec{\cancel{E}}_T)^2}.
\end{equation}
and a cut of 300 GeV on this quantity 
proves to be effective in further reducing the background.
On the hadronic side,
the 4-jet invariant mass should peak at the location of $M_{H^{++}}$.
Taking into account the detector resolution, we look for the events in
the window $M_{H^{++}} \pm 50\ (30)$ GeV, which reduces the background by  
another order of magnitude. 
Two sets of numerical values for the cuts have been presented in
Table \ref{Tab:I}, to establish the fact that the 
successive criteria applied here can reduce the backgrounds
by more than two orders of magnitude, 
while the survival efficiency for the signal 
remains as high as $35\%$.
This example illustrates how the rather distinctive 
kinematical features of the signal allow us to effectively suppress the
SM backgrounds, after which the signal observation is 
mainly a statistical issue.


\begin{table}[tb]
\begin{tabular}{| c || c | c |c | c| c | c| c |}
    \hline
     (fb) & Basic Cuts  & $p_T^\ell$ cut &  $p_T^j$ cut & $M_W$ reconst. & $M_{\rm Cluster}$   & $M_T$ & $M_{jjjj}$ \\
\hline
\hline
cuts & $\cancel{E}_T>40~{\rm GeV}$ & $>70~{\rm GeV}$ & $>120~{\rm GeV}$ & $M_W\pm 15~{\rm GeV}$ & $>600~{\rm GeV}$ & $>300~{\rm GeV}$ & $300\pm 50~{\rm GeV}$\\ 
     \hline
      signal & 0.20  & 0.17 & 0.15 & 0.13 & 0.13 & 0.12 & 0.12 \\
\hline
     $t\bar{t} W$ & 4.15  & 2.52 & 1.60 & 0.82 & 0.74 & 0.48 & 0.12\\
   \hline
\hline
cuts & $\cancel{E}_T> 70~{\rm GeV}$ &  $>120~{\rm GeV}$ & $>120{\rm GeV}$ & $M_W\pm 15~{\rm GeV}$ & $>600~{\rm GeV}$ & $<300~{\rm GeV}$ & $300\pm 30{\rm GeV}$\\
\hline
signal & 0.17 & $9.71\times 10^{-2}$ & $9.08\times 10^{-2}$ &  $8.27\times 10^{-2}$ &  $8.24\times 10^{-2}$ & $7.80\times 10^{-2}$ & $6.18\times 10^{-2}$\\
\hline
$t\bar{t}W$ &  2.82 & 0.79 & 0.62 & 0.32 & 0.30 & 0.15 & $2.90\times 10^{-2}$\\
\hline
  \end{tabular} 
  \caption{The signal and the leading background rates for $pp\rightarrow W^+ W^+ W^- W^-\rightarrow jjjj+\ell^+\ell^+ +\cancel{E}_T$. The rates after imposing each selection criterion, as described in
the text, are shown. $m_{H^{++}}=300\ {\rm GeV}$. $\mu_R =\mu_F = (2m_t + M_W)/2$ for 
$t\bar{t}W$ calculation.}
\label{Tab:I}
\end{table}


If the doubly charged Higgs is heavier, the $W$'s arising from its decay 
will be boosted, so that the two jets from each of the hadronically 
decaying ones  are rather highly collimated. 
The typical opening angle of each pair from $W$-decay resulting into a single fat
jet of this kind is 
$2M_W/M_{H^{++}}\approx 0.27$ for $M_{H^{++}}=600$ GeV,
and the peak in $\Delta R$ bwteen jets coming from the hadronically
decaying $W$'s  occurs around 0.4. This means that a substantial 
fraction of the four parton-level jets for such a high Higgs mass
will actually be merged into two-jet events, together with the like-sign dileptons.
Such final states  are apparently threatened by the overwhelming 2-jet backgrounds. 
In order to get rid of those backgrounds, one can first demand high transverse energy  
($E_T^J$) for both of the jets, as shown in the second column of  Table \ref{Tab:II}. 
In addition, one can utilize the ``fatness" of the jets 
as a discriminator. We define a fat jet ($J$) with cone size
\begin{equation}
\Delta R_{J} < 0.8 
\end{equation}
%
Since the two jets merged into one $J$ in the case of the signal will have an invariant mass 
peaking at $M_W$, we  further demand two such mass peaks, each within 
\begin{equation}
m_{J_1} = m_{J_2} = M_W \pm\ 15\ {\rm GeV}.
\end{equation}
While all signal events  show this feature,  
the typical  jets  from QCD partons would not have large mass.
In reality, however, light quarks and gluons do develop parton showers
due to QCD radiation and thus lead to finite mass. 
Some simulations show that an effective jet mass scales with its transverse
energy roughly like $m_J \approx (10-15)\%  E_T^J$. For a jet to acquire  a mass 
of the order of $M_W$, say 65 GeV, its transverse energy would have to be
at least 500 GeV. We will thus require an $E_T^J$ on the order of 500 GeV
for the background jets. The demand of
such a high  $E_T^J$  suppresses the backgrounds quite effectively.
The final killer of the backgrounds is the demand to reconstruct 
$M_{H^{++}}=m(W^+W^+)=M_{JJ}$. 

We illustrate the effectiveness of these
steps by explicitly evaluating a leading background $jjW^\pm W^\pm$ from
QCD processes. The results are listed in Table II. Once again,
the signal observation is mainly a statistical issue after the elimination of the
backgrounds. As already seen in Fig.~\ref{ww}(b), 
the surviving rates persented here show that 
an integrated luminosity of 300 fb$^{-1}$  should allow for extracting
the signal of about 700 GeV doublly charged Higgs in this channel. 

\begin{table}[t]
\begin{tabular}{| c || c | c| c | c|}
    \hline
  Rate & Basic Cuts  & ${\rm max}(E_T^J)> 200~{\rm GeV}$ &  Jet mass $m_J$  & $M_{JJ}$ \\

 (fb)& & ${\rm min}(E_T^J)>140~{\rm GeV}$  & $M_W\pm 15~{\rm GeV}$ & $600\pm 75~{\rm GeV}$\\ 
     \hline
      signal & $3.62\times 10^{-2}$  & $3.61\times 10^{-2}$ & $ 3.60\times 10^{-2}$ & $3.60\times 10^{-2}$ \\
\hline
     $JJ W^\pm W^\pm$ & 14.53  & 4.66 & 0.18 &  $1.68\times 10^{-4}$ \\
   \hline
  \end{tabular}
  \caption{The signal and the leading background rates for $pp\rightarrow W^+ W^+ W^- W^-\rightarrow JJ+\ell^+\ell^+ +\cancel{E}_T$. The rates after imposing each selection criterion, as described in
the text, are shown. $m_{H^{++}}=600\ {\rm GeV}$. $\mu_R =\mu_F = \sqrt{s/4}$ 
for $JJW^\pm W^\pm$ calculation.}
\label{Tab:II}
\end{table}

\section{Summary and conclusions}
We have investigated the visibility of the pair-production of doubly charged scalars
at the LHC. Such a scalar is assumed to belong to an SU(2)$_L$ triplet which can generate
Majorana masses for left-handed neutrinos through the vev of its neutral component.
Pair-production,  in spite of its relative 
kinematical suppression,  has the advantage of being relatively
model-independent, and is not driven by the vev of the triplet. We find that, while
the contribution comes largely from the Drell-Yan process, two-photon fusion also
contributes at the level of 10\% at the LHC due to the substantially enhanced
electromagnetic coupling. This, we emphasize, is comparable to
the QCD correction to the Drell-Yan channel, and must be included in a complete and
 accurate estimate. 

Signatures of the pair-produced scalars have been investigated over the entire 
parameter space where viable neutrino masses can arise from the scalar triplet.  
We have covered both the regions where they decay dominantly into 
like-sign dileptons and like-sign $W$'s. The former are largely free from backgrounds,
and a mass  range upto about $800~{\rm GeV}$ to $1~{\rm TeV}$ can be probed 
with an integrated luminosity 
of 300 ${\rm fb}^{-1}$,  if the doubly charged Higgs has a branching ratio 
of at least $30\% - 60\% $  in this channel. In the latter channel, we have 
suggested optimal ways of eliminating the standard model backgrounds
with high efficiency for the signal retention.  
The signal observation is mainly a statistical issue,
and a doubly charged scalar of mass up to about 700 GeV should be identifiable, 
if it dominantly decays into a pair of like-sign $W$'s.
With the spirit of neutrino mass generation of Eq.~(2) and the signal
complementarity between the $\ell^{+}\ell^{+}$ and $W^+W^+$ channels
as shown in Fig.~\ref{ww}(b) by the 3-event contours, 
we claim a complete coverage for $M_{H^{++}} \approx 650$ GeV with any 
arbitrary decay of these two modes.

There has been great interest of searches for long-lived doubly charged scalars.
We found that in order to have the requisite contribution to neutrino masses,
it is not possible for the scalar to be long-lived, although it is possible to leave 
a perceptible decay gap at the detector for  $M_{H^{++}} \lsim 250$ GeV.
Any observation of long-lived scalars of this kind will
therefore be a pointer towards some alternative theory of neutrino 
masses such as the Type II seesaw mechanism.  Finally, we wish to emphasize that
a study of the decays of the doubly charged scalar into different lepton flavors,
diagonal as well as off-diagonal,  is a very useful probe of the mechanism whereby
neutrino masses arise from interaction with an SU(2)$_L$ scalar triplet.  This probe is
eminently  within the scope of the LHC. 
 
\section*{Acknowledgement}
We thank Patha Konar for helpful discussions. The work of TH and 
KW is supported in part by the U.S. Department of Energy under 
grant No. DE-FG02-95ER40896, by the Wisconsin Alumni Research Foundation. 
BM would like to thank the Department of Atomic Energy, India, for support 
through the Xth and XIth 5-Year Plan Projects. 
Si is supported by NNSFC, NCET and Huoyingdong Foundation. 
BM and Si would like to acknowledge the hospitality of the 
Phenomenology Institute, University of Wisconsin-Madison while 
the work was in progress and completed.

\end{document}